\documentstyle[preprint,aps]{revtex}
\begin{document}
\draft \title{Pressure Dependence of Born Effective Charges,
  Dielectric Constant and Lattice Dynamics in SiC} \author{Cheng-Zhang
  Wang, Rici Yu and Henry Krakauer} \address{Department of Physics,
  College of William and Mary, Williamsburg, Virginia 23187-8795}
\date{August 14, 1995} \maketitle

\begin{abstract}
  The pressure dependence of the Born effective charge, dielectric
  constant and zone-center LO and TO phonons have been determined for
  $3C$-SiC by a linear response method based on the linearized
  augmented plane wave calculations within the local density
  approximation.  The Born effective charges are found to increase
  nearly linearly with decreasing volume down to the smallest volume
  studied, $V/V_0=0.78$, corresponding to a pressure of about 0.8
  Mbar.  This seems to be in contradiction with the conclusion of the
  turnover behavior recently reported by Liu and Vohra [Phys.\ Rev.\
  Lett.\ {\bf 72}, 4105 (1994)] for $6H$-SiC.  Reanalyzing their
  procedure to extract the pressure dependence of the Born effective
  charges, we suggest that the turnover behavior they obtained is due
  to approximations in the assumed pressure dependence of the
  dielectric constant $\varepsilon_\infty$, the use of a singular set
  of experimental data for the equation of state, and the uncertainty
  in measured phonon frequencies, especially at high pressure.

  \vskip 0.2truein\noindent {\sl Submitted to Phys.\ Rev.\ B1}
\end{abstract}

\newpage \narrowtext

\section{Introduction}

Recently, Liu and Vohra\cite{liuvohra} presented intriguing evidence
regarding the pressure dependence of the Born effective charge in
$6H$-SiC.  These authors found that the effective charge increased
initially with increasing pressure, reaching a maximum at about 0.4
Mbar. Further increasing the pressure, however, resulted in a decrease
in the magnitude of the effective charge. Because of its great
potential in device applications, especially in harsh environments, it
is important to accurately characterize the properties of this unusual
material. SiC crystallizes in hundreds of polytypes, corresponding to
different stacking sequences of Si-C bilayers in the cubic \mbox{[1 1
  1]} direction.\cite{verma} Polytypes with cubic, hexagonal and
rhombohedral symmetry are designated $3C$, $nH$ and $nR$ respectively,
where $n$ is the number of layers in the repeating unit, with the $3C$
structure corresponding to the zincblende structure.  The different
polytypes have very similar properties, since differences in local
atomic coordination first appear in the second neighbor shell and the
experimental equation of state for the different polytypes are very
similar.\cite{goncharov90,yoshida93} Phonon dispersion along the
stacking direction is considered universal for different prototypes
and this universality has been used to map out the phonon frequencies
in this direction from Raman scattering measurements.\cite{feldman68}
Indeed, a recent self-consistent study of $3C$, $2H$ and $4H$ SiC by
Karch {\it et al.}\cite{karch94} found only small differences in the
calculated phonon frequencies along the stacking direction.  The main
purpose of the present work is to report a first-principles study of
structural and dynamical properties of SiC under high pressure.  In
particular, the volume dependence of the Born effective charges and
the dielectric constant were obtained.  We find that the Born
effective charge increases nearly linearly as the volume is decreased.
This is in sharp contrast with the above-mentioned turn-over behavior
reported recently for $6H$-SiC.  The likely reasons for this
discrepancy will be discussed in light of the calculated results.

\section{Methodology}

First principles total-energy calculations in the local-density
approximation (LDA) were performed using the linearized augmented
plane wave (LAPW) method\cite{lapw75-2} to determine the ground state
structural properties of $3C$-SiC at equilibrium and under pressure.
Lattice dynamical properties, Born effective charges, and the
dielectric constant were obtained with a linear response algorithm
developed recently within the LAPW method.\cite{yu94,wang94,yu95} To
dispense with the need to treat the chemically inert localized inner
core orbitals, we employ a hard Kerker\cite{kerker} type
pseudopotential. The muffin-tin radii for Si and C are 1.79 and 1.50
$a.u.$, respectively. We used the Wigner interpolation formula for
exchange-correlation potential.\cite{wigner} Born effective charges
and the dielectric constant were calculated using up to a \mbox{
  $12\times 12\times 12$} uniform {\it k}-point mesh\cite{monk}, which
corresponds to 28 special points in the irreducible Brillouin zone
(BZ). At this level of convergence, the acoustic sum-rule is satisfied
with an error of only 0.6\%. In calculating the dynamical matrices, we
used an \mbox{ $8\times 8\times 8$} {\it k}-point mesh. Tests with
denser meshes show that phonon frequencies are converged to better
than about 0.5\%. Tests with other exchange-correlation functionals
also display differences at this level. Phonon dispersions in the
harmonic approximation were obtained in the full BZ as follows.
First, {\it ab initio} calculations were carried out to determine the
dynamical matrix at the irreducible phonon wavevectors corresponding
to a \mbox{ $4\times 4\times 4$} uniform mesh, which by symmetry gives
the dynamical matrix at all mesh points. The dynamical matrix can then
be obtained at arbitrary wavevectors by interpolation, first
separating it into a long-range dipole-dipole term and a short-range
term.\cite{gonze94} The former is evaluated exactly from the
calculated Born effective charges and dielectric constant using the
Ewald summation technique.  The remaining short-range part is then
interpolated using real-space force constants, which are found through
Fourier transform.

\section{Results and Discussion}

\subsection{Structural Properties, Equation of State and Phase Transition}

LAPW total energy calculations were performed to determine the
theoretical lattice constant $a_0$, bulk modulus $B_0$ and its
pressure derivative $B^{'}_{0}$ for $3C$-SiC by fitting to the
Murnaghan equation of state.\cite{murnaghan} Earlier experimental and
theoretical studies suggest that the pressure-volume relations for
different polytypes are similar, due to the rigidity of the
nearest-neighbor coordination.  The equation-of-state data for the two
polytypes $3C$-SiC and $6H$-SiC have been found to be essentially the
same up to the transition pressure of $3C$-SiC.\cite{yoshida93}
Theoretical calculations\cite{karch94} also yield very small
difference in the bulk modulus (less than 1\%) between the $3C$, $2H$
and $4H$ polytypes.  The calculated pressure derivative of the bulk
modulus for these polytypes are different by only 7\%.\cite{karch94}

Results for $3C$-SiC are presented in Table~\ref{equil-sic} and
compared with other calculations and experiment for $3C$ and other
polytypes.  Our results agree well with other calculations given in
the table.  Slight differences may be due to the use of different
forms of the exchange-correlation potential. There is nearly perfect
agreement between the present calculations and those of Chang and
Cohen\cite{chang87}, both using the Wigner exchange-correlation
potential.  Karch {\it et al.}\cite{karch94} used the Ceperly and
Alder exchange-correlation formula\cite{ceperl} as parametrized by
Perdew and Zunger.\cite{perdew} Lambrecht {\it et al.}
\cite{lambrecht91} used an LMTO basis set and the von Barth-Hedin
parametrization of the exchange-correlation energy.\cite{barth} As
also seen in other systems\cite{lu-dis}, the Wigner form tends to
yield slightly larger equilibrium lattice constants than other forms,
and as expected the bulk modulus is accordingly slightly smaller.

A number of high-pressure experiments have been carried out on various
polytypes of SiC.  While they generally show little difference between
the equation of state for different polytypes, there are large
discrepancies between the results of different experimental groups for
the bulk modulus and its pressure derivative.  Most earlier
experiments report a bulk modulus around 2.25 Mbar.  The recent work
of Yoshida {\it et al.}\cite{yoshida93} on SiC polytypes up to 0.95
Mbar, the highest pressurization on these materials so far, reported a
relatively large bulk modulus $B_0$ ($2.60\pm0.09$~Mar) and a smaller
pressure derivative $B^{'}_{0}$ ($2.9\pm0.3$).  They ascribed the
discrepancy with the previous measurements to fitting to the larger
pressure range in their experiment.  Analysis of our calculated
results does not support this explanation.  Using our calculated
total-energy at various volumes, we can fit to the Murnaghan equation
of state\cite{murnaghan} for pressures up to about 0.8 Mbar.  We find
only small differences (both $B_0$ and $B^{'}_0$ change about 1 $\%$)
compared to fitting over a smaller pressure range.  The explanation is
also inconsistent with the experiments of Aleksandrov {\it et
  al.}\cite{aleksandrov} and Goncharov {\it et al.}\cite{goncharov90},
which were carried out respectively to 0.425 Mbar and 0.45 Mbar but
obtained smaller bulk modulus as compared with that of Str\"ossner
{\it et al.}\cite{strossner87}, which went up to only 0.25 Mbar.
Based on the results of all existing well-converged theoretical
calculations, it seems reasonable to suggest that the lower range of
the experimental bulk modulus is more likely to be correct.

We have also determined the volume dependence of the total energy for
rocksalt structure SiC. Table~\ref{trans-sic} compares the calculated
transition parameters to other calculations and experiment.  The
agreement between the different calculations is good, predicting that
the transformation from the zincblende phase to the rocksalt phase
takes place at about 0.65 Mbar. The volume of the zincblende phase at
the phase transition is predicted to be about 82\% of its equilibrium
volume at ambient pressure, and the volume reduction accompanied by
the transition is predicted to be about 20\%.  The predicted
transition pressure and the volume of the $3C$-SiC phase before the
transition differ considerably from the experimental data, but the
observed volume reduction is in good agreement.  This may be due to
the fact that the experimental transition pressure was obtained for
the forward transition (from zincblende phase to rocksalt phase),
where an excess pressure beyond the equilibrium value appears and is
included in the measurements.\cite{yoshida93}

\subsection{Born Effective Charge, Dielectric Constant, Phonon
  Frequencies, and Elastic Constants at Equilibrium Volume}

The Born effective charge $Z^{*}$ and dielectric constant
$\epsilon_{\infty}$ are calculated at small {\bf q} wavevector \mbox{
  0.01~[1~1~1]$\frac{2\pi}{a}$}, where $a$ is the lattice constant.
Table~\ref{z-sic} presents our results at the experimental equilibrium
volume.  The Born effective charges $Z^{*}$(Si) and $Z^{*}$(C) are
converged to better than one percent when BZ integrations are
performed using a 28 special $k$-point set in the irreducible BZ,
corresponding to a \mbox{$12\times 12\times 12$} uniform $k$-point
sampling.  The acoustic sum rule requires that $Z^{*}$(Si) and
$Z^{*}$(C) have the same magnitude and opposite sign.  Violations
arise from finite $k$-point sampling. However, good results can be
obtained with 10 special $k$ points, especially after averaging the
magnitudes of $Z^{*}$(Si) and $Z^{*}$(C).\cite{giann91} The averaging
also reduces the effect of using a small but finite wavevector.  This
effect was checked by using a slightly larger wavevector \mbox{ {\bf
    q}=0.02~[1~1~1]$\frac{2\pi}{a}$}, which yielded only a slightly
larger discrepancy.  Our calculations give a Born effective charge in
excellent agreement with experiment and with the linear response
calculations by Karch {\it et al.}, using a plane wave basis
set\cite{karch94}. The static dielectric constant $\epsilon_{\infty}$
is about 7\% too large, however. The tendency of the LDA to
overestimate $\epsilon_{\infty}$ is a well known problem that is
attributed not to the density functional theory, but to the LDA used
for the exchange-correlation potential.\cite{corso94,gonze95}

Our calculated phonon frequencies are shown in Fig.~\ref{dispersion}.
For points along the \mbox{ $\Lambda[\xi~ \xi~ \xi]$} direction, we
obtained the dynamical matrices as described in Section II, using a
uniform \mbox{$4\times 4\times 4$} phonon wavevector mesh.  But along
the \mbox{$\Delta[\xi~ 0~ 0]$} and \mbox{$\Sigma[\xi~ \xi~ 0]$}
directions, we have calculated the dynamical matrices at additional
points, which correspond to an \mbox{$8\times 8\times 8$} mesh, and
obtained one-dimensional interplanar force constants to perform the
interpolation.\cite{sriva-b-90} This improves the interpolated
dispersion in the acoustic region. No neutron scattering data for
zincblende SiC is available because it is difficult to grow large
enough $3C$-SiC single crystals.  The filled squares in the figure are
experimental phonon frequency data from Raman measurements.  The
experimental results on the BZ boundary (X and L) are obtained from
second-order Raman scattering\cite{olego82-2}.  Others are first-order
Raman data from hexagonal and rhombohedral polytypes that have been
unfolded into the larger $3C$-SiC BZ.\cite{feldman68} The excellent
agreement shows that the different stacking sequences have little
influence on the vibrational properties. As mentioned, the calculated
phonon frequencies of Karch {\it et al.}\cite{karch94} on $2H$- and
$4H$-SiC directly support this conclusion.  Table~\ref{freq} compares
our calculated phonon frequencies at high-symmetry {\it k}-points with
those of Karch {\it et al.}\cite{karch94} and with experiment.  The
zone-center $\omega_{\rm LO}$ frequency is obtained from the
calculated $\omega_{\rm TO}$, $\epsilon_{\infty}$, $Z^{*}$, and the
volume of the primitive unit cell $V$, using the relation
\begin{equation}
  \omega^2_{\rm LO} = \omega^2_{\rm TO } + { { 4 \pi e^2{\bf Z}^{*2} }
    \over {\varepsilon_\infty}\mu V } ,
\label{loto}
\end{equation}
where $\mu$ is the reduced mass.  Our calculated frequencies are
systematically about 1\% smaller than those of Ref.  \cite{karch94}.
This is due in part to the difference in the equilibrium volume in the
two calculations.

The elastic constants can be extracted from the calculated acoustic
phonon dispersions.  For the cubic structure, the elastic constants
$C_{44}$ and $C_{11}$ are simply related to the velocities of
transverse and longitudinal plane waves in \mbox{[1 0 0]} direction as
follows\cite{sriva-b-90}:
\begin{equation}
  \upsilon_{T} [100] = \sqrt {\frac{C_{44}}{\rho} },
\end{equation}
\begin{equation}
  \upsilon_{L} [100] = \sqrt {\frac{C_{11}}{\rho} },
\end{equation}
where $\rho$ is the mass density of SiC.  The velocities of $
\upsilon_{T}$[100] and $\upsilon_{L}$[100] can be obtained from the
slopes of the acoustic transverse and longitudinal phonon dispersions
at the zone center in the [1 0 0] direction:
\begin{equation}
  \upsilon_j[100] = \frac{\partial \omega_j} {\partial k},
\end{equation}
where $\omega_{j}$ is the angular frequency of the $j$-th branch
phonon and $k$ is the magnitude of the corresponding wavevector.
Similarly, the elastic constant $C_{12}$ can be obtained from the
relationship \cite{sriva-b-90}
\begin{equation}
  \upsilon_{T_1} [110] = \sqrt {\frac{C_{11}-C_{12}}{2\rho} },
\end{equation}
where $\upsilon_{T_1}$ is the velocity of transverse wave with atomic
displacement along \mbox{[1 $\bar{1}$ 0]}, which corresponds to the
lowest acoustic branch in the $\Sigma$ direction, i.e. \mbox{[1 1 0]}.
The bulk modulus $B_{0}$ is related to these elastic constants by the
relation\cite{cochran73}
\begin{equation}
  B_{0}=\frac{ C_{11} + 2 C_{12} } {3}.
\end{equation}
The directional dependence of the transverse sound velocity in
$3C$-SiC is related to the fact that the interatomic forces in
$3C$-SiC are highly noncentral, which implies that the elastic
constants do not satisfy the Cauchy relation ($C_{12}=C_{44}$).
Table~\ref{elastic-sic} presents the elastic constants and the bulk
modulus as extracted using the equations above, together with other
calculated results and experiment.  Our results are in good agreement
with experiment and slightly smaller than the calculations by Karch
{\it et al.} \cite{karch94} due again in part to the use of our
slightly larger lattice constant.  The bulk modulus $B_{0}$ determined
from the elastic constants is consistent with our total energy
calculations in the previous section.

\subsection{Pressure Dependence of Phonon Frequencies,
  Dielectric Constant, and Born Effective Charge}

The mode Gr\"uneisen parameters, defined as
\begin{equation}
  \gamma = - \frac{V}{\omega} \frac{ \partial \omega} { \partial V},
\end{equation}
describes the volume dependence of the phonon frequencies.  The
calculated mode Gr\"uneisen parameters at the $\Gamma$, $X$ and $L$
points in the BZ are presented in Table~\ref{grun}. Our results appear
to be in good agreement with Karch {\it et al.}\cite{karch94} as
presented in their Fig. 9.  In the text of that paper, they also gave
the numerical value for the TA($X$) mode, 0.13, which compares
favorably with our value of 0.12.  The experimental results of Olego
{\it et al.}\cite{olego82-2,olego82} and of Aleksandrov {\it et
  al.}\cite{aleksandrov} are presented for comparison.  In the former
work, the Gr\"uneisen parameters were calculated from their
experimental volume dependence using a bulk modulus of 3.219 Mbar.
This value is likely to be too large, as discussed above.  The values
given in Table~\ref{grun} are converted from their results, assuming
the bulk modulus from our calculations.  The results are seen to be in
generally good agreement with our calculation.  A significant
discrepancy exists for the LA mode at the $L$-point.  The experiment
of Olego {\it et al.} gave a negative value, in contrast with our
theoretical prediction and that of Karch {\it et al.}\cite{karch94}

Fig.~\ref{fre-lto-pressure} compares the calculated pressure
dependence of the zone-center phonon frequencies of 3$C$-SiC with the
experiments for both 3$C$-SiC and 6$H$-SiC. The numerical values of
our frequencies are presented in Table~\ref{zvolt}, together with the
calculated Born effective charge and dielectric constant of $3C$-SiC
at different lattice constants.  The second row of Table~\ref{zvolt}
gives the results at the equilibrium volume and the smallest lattice
parameter corresponds to a pressure of about 0.8 Mbar.  The calculated
pressure as a function of volume was determined by fitting the volume
dependence of the total energy to the Murnaghan equation of
state\cite{murnaghan}.  We have also tested fitting the
Birch-Murnaghan equation of state\cite{birch78}, which was used by
Yoshida {\it et al.}\cite{yoshida93}, and find only small differences,
with $B_0$ increasing by 1$\%$, $B^{'}_0$ increasing by less than
3$\%$, and the pressure values changing less than 1$\%$.  As in the
ambient pressure case mentioned above, our phonon frequencies are
nearly uniformly a few percent smaller than those of 3$C$-SiC measured
up to 0.25 Mbar by Olego {\it et al.}\cite{olego82}.  Therefore the
experimental pressure dependence of phonon frequencies of zone-center
LO and TO modes are fairly well reproduced in our calculations. This
is consistent with the agreement of zone-center mode Gr\"uneisen
parameters in Table~\ref{grun} between theory and experiment.
Fig.~\ref{fre-lto-pressure} also compares our calculations with Liu
and Vohra's measurements for 6$H$-SiC.  With increasing pressure, our
calculated frequencies stiffen slightly faster than their experiment,
with values initially about 2$\%$ smaller than measurements changing
to 2$\%$ larger at the end. To see the differences more clearly, the
resulting LO-TO splitting from theory and experiments is presented in
Fig.~\ref{lto-split}.  For $3C$-SiC, Olego {\it et al.}'s measured
results under pressure\cite{olego82} appear to increase faster than
our calculations.  In 6$H$-SiC, Liu and Vohra\cite{liuvohra} observed
a rapidly increasing LO-TO splitting at low pressure which saturates
at the high end of the pressure range studied (0.9~Mbar).  Our
calculations compare well with the trend of the pressure dependence in
their measurements, but the calculated values are smaller due to the
overestimation in the dielectric constant.  The pressure dependence of
this overestimation is unknown, of course.  However, if we assume that
this is not a significant factor and reduce the calculated
$\varepsilon_{\infty}$ by 7\%, the calculated LO-TO splitting
increases as shown by the dotted lines in Fig.~\ref{lto-split}, which
is in good agreement with the result of Liu and Vohra, except at the
highest pressures.

As mentioned, using Eq.~(\ref{loto}), Liu and Vohra\cite{liuvohra}
deduced the pressure dependence of the Born effective charges from the
LO-TO splitting.  Their results show that the effective charge
increases at low pressure but reaches a maximum at about 0.4~Mbar and
decreases as the pressure is increased further.  Actually, the Born
effective charges in the hexagonal polytypes are slightly anisotropic
and are different for atomic displacement in the planes and
perpendicular to the planes.  The difference is about 7\% for $2H$ and
$4H$ polytypes.\cite{karch94} This anisotropy in the effective charge
and other physical quantities (such as the dielectric constant) was
neglected in the analysis of Liu and Vohra.\cite{liuvohra} If we
consider the averaged effective charge, the results of Karch {\it et
  al.}\cite{karch94} show that there is little distinction between the
cubic and the hexagonal polytypes.  In this sense, the results of our
calculation for the cubic SiC, discussed below, are pertinent to the
above-mentioned experiment.

In order to extract the pressure dependence of the effective charge
from the measured LO-TO splitting, two additional pieces of
information are required: i) the equation of state to derive the
volume from the measured pressure, and ii) the volume dependence of
$\varepsilon_\infty$.  By contrast, our calculated Born effective
charge is determined directly and is known to be very accurate, unlike
the dielectric constant which suffers from the LDA approximation.  For
their equation of state, Liu and Vohra used the values of Yoshida {\it
  et al.}\cite{yoshida93} for the bulk modulus, $B = 2.60$ Mbar, and
its pressure derivative, $B^\prime = 2.9$ together with the
Birch-Murnaghan equation of state.  However, this bulk modulus is 10 -
20\% larger than the other measured values, and the pressure
derivative is about 20 - 35\% smaller.

The volume dependence of $\varepsilon_\infty$ has apparently never
been measured for SiC.  Liu and Vohra followed Olego {\it et
  al.}\cite{olego82} in using the following expression,
\begin{equation}
  {{\partial \ln \varepsilon_\infty}\over{\partial \ln V}} = r ,
\label{dedv}
\end{equation}
with $r = 0.6$ for this range of pressures, citing earlier work on
both Si and C that reported this value. Go\~{n}i {\it et
  al.}\cite{goni} have reviewed the volume dependence of the
refractive index of Ge and GaAs. Their Table II shows large
differences between various measurements of the volume derivative of
$\varepsilon_{\infty}$ in Eq.  (\ref{dedv}). Thus, the use of the
scaling relation, Eq. (\ref{dedv}), and the particular value $r = 0.6$
must be regarded as a weak link in the experimental analysis of the
volume dependence of the Born effective charge.  Indeed, the present
calculations suggest that this value of the exponent scaling power is
too large.  In Fig.~\ref{dedvfig}, we compare the scaling relation Eq.
(\ref{dedv}) with the calculated volume dependence of
$\varepsilon_\infty$.  The value $r \simeq 0.3$ is found to give the
best fit to the calculated values.  As mentioned, the LDA tends to
overestimate $\varepsilon_\infty$, in this case by about 7\%. The
volume dependence of this error is unknown, but it would have to be
large to modify the above conclusions.

In Fig.~\ref{zvol}, we compare the volume dependence of the Born
effective charge $Z^{*}$ from calculations and those derived from
experiment. The calculated Born effective charge for 3$C$-SiC is seen
to linearly increase with decreasing volume (see curve ($a$)), and
there is no evidence of a relative maximum as is in the experimental
results of Liu and Vohra for 6$H$-SiC\cite{liuvohra} .  The other four
curves are all derived from experiment, using Liu and Vohra's
quadratic fit to the measured LO-TO splitting for 6$H$-SiC, but with
different choices of equation of state and volume dependence of
dielectric constant.  In curve $(b)$, we have used the same input data
as Liu and Vohra\cite{liuvohra}: the scaling relation Eq. (\ref{dedv})
with $r=0.6$ (using the experimental $\varepsilon_\infty = 6.52$ at
ambient pressure) and the equation of state ($B = 2.60$ Mbar and
$B^\prime = 2.9$) from Yoshida {\it et al.}\cite{yoshida93}.  This
curve exhibits the same peaked behavior as that shown in Fig. 3 of Liu
and Vohra\cite{liuvohra}.  Curve $(c)$, which is plotted with the same
input data as for curve $(b)$ except using our calculated equation of
state ($B = 2.20$ Mbar and $B^\prime = 3.71$), is much less peaked
than curve $(b)$, indicating that the volume dependence of $Z^*$
derived from experiment following Eq.~(\ref{loto}) is quite sensitive
to the equation of state.  The other two curves $(d)$ and $(e)$ are
derived using the best scaling relation Eq. (\ref{dedv}) with $r=0.3$
(again using the experimental $\varepsilon_\infty = 6.52$ at ambient
pressure).  We have used the equation of state from Yoshida {\it et
  al.} for curve $(d)$ and our calculated equation of state (for
3$C$-SiC) for curve $(e)$.  These two curves give larger Born
effective charges than $(b)$ and $(c)$ since the scaling relation $
r~=~0.3$ yields larger dielectric constant under pressure than
$r~=~0.6$.  Curve $(e)$ is seen to agree quite well with the
calculated $Z^*$ except for volumes smaller than about 0.85 $V_0$.
Examination of Liu and Vohra's measurements in Fig.~\ref{lto-split},
which shows the scatter in the experimental values of the LO-TO
splitting, indicates that the greatest uncertainties occur for the
higher pressures (smaller volumes) due to the pressure induced
broadness of Raman and ruby peaks\cite{liu-private}.  The magnitude of
this scatter $\simeq 4 cm^{-1}$ (or 2$\%$) taken together with the
questionable use of the scaling relation Eq.  (\ref{dedv}) with
$r=0.6$ and the large bulk modulus of Yoshida {\it et al.} undermines
the conclusions of Liu and Vohra\cite{liuvohra} regarding the peaked
behavior of $Z^*$. Given the uncertainties in the experimental
analysis, we would assert that our prediction of a linear increase of
$Z^*$ with decreasing volume, up to pressures of 0.8 Mbar is not
contradicted by their measurements.

\section{Conclusions}

We have presented a detailed analysis of the volume dependence of
various properties of 3$C$-SiC.  The calculated equilibrium lattice
constant, bulk modulus and pressure derivative of the bulk modulus of
3$C$-SiC were found to agree well with experiment and other LDA
calculations.  The transition pressure from the zincblende phase to
the rocksalt phase is determined to be 0.65 Mbar, in good agreement
with other theoretical results but lower than the reported
experimental value (1.00 Mbar). The volume reduction, however, is in
good agreement.  The discrepancy may be due in part to the fact that
the experimental transition pressure obtained in a diamond-anvil
apparatus is for the forward transition (from zincblende phase to
rocksalt phase), which requires an excess pressure\cite{yoshida93}
beyond the equilibrium value.  The equilibrium calculated Born
effective charge $Z^{*}$ is in excellent agreement with experiment,
but the dielectric constant is about 7\% too large, as is typical in
LDA calculations.  The phonon dispersion and the elastic constants
agree with available experiment data, demonstrating that the different
stacking sequences in silicon carbides have little influence on the
vibrational properties.  The experimental Gr\"uneisen parameters are
well reproduced, and predictions are given for those experimentally
unavailable.  In $3C$-SiC, we find that i) the dielectric constant
$\varepsilon_\infty$ decreases with increasing pressure, and ii) the
Born effective charge increases monotonically with pressure, without
any evidence of a relative maximum as reported for $6H$-SiC by Liu and
Vohra \cite{liuvohra}.  After analyzing the procedure used to extract
the experimental Born effective charges $Z^{*}$ of $6H$-SiC in
Ref.\cite{liuvohra}, we suggest that the turn-over behavior of $Z^{*}$
reported by Ref.\cite{liuvohra} is due to the assumptions regarding
the volume dependence of dielectric constant $\varepsilon_\infty$, the
use of a singular set of experiment data for $B_{0}$ and $B^{'}_{0}$,
and uncertainties in the measured phonon frequencies, especially at
high pressure.  Our calculations predict a linear increase of $Z^*$
with decreasing volume, up to pressures of 0.8~Mbar.

{\it Note Added.} Since the submission of the manuscript, two recent
publications\cite{tang95,sengstag95} have come to our attention.
A theoretical study using molecular dynamics simulations with an empirical
potential model\cite{tang95} raised the possibility of
amorphization of SiC under high pressure.  This seems to be inconsistent
with the experiment of Yoshida {\it et al.},\cite{yoshida93} which observed
a direct polymorphic transition to the rocksalt structure at 1~Mbar. Thus
further experimental and theoretical work needs to be done to see if this is a
real effect. Ref.\cite{sengstag95} presented calculated Born effective charges
of 3$C$-SiC under low pressure, in good agreement with this work

\acknowledgments We acknowledge useful discussions with W. E. Pickett,
D. Singh and G. Gu. We thank J. Liu and Y. K. Vohra for providing
original experimental data on 6$H$-SiC.  Supported by Office of Naval
Research grant N00014-94-1-1044. C.-Z. Wang was supported by National
Science Foundation Grant DMR-9404954. Computations were carried out at
the Cornell Theory Center.

\newpage

\clearpage
\begin{table}
\caption{Equilibrium state properties of SiC. The equilibrium lattice
  constant, bulk modulus and its pressure derivative are represented
  by $a_0$, $B_0$, and $B^{'}_0$, respectively.}
\label{equil-sic}
\begin{center}
\begin{tabular}{lllcc}
  & $a$ (\AA) & $B_{0}$ (Mbar) & $B^{'}_{0}$ & Pressure range (Mbar)
  \\ \hline Theory & & & & \\ Chang {\em et al.}$^{a}$ & 4.361 & 2.12
  & 3.7 & \\ Lambrecht {\em et al.}$^{b}$ & 4.315 & 2.23 & 3.8 & \\
  Karch {\em et al.}$^{c}$ & 4.345 & 2.22 & 3.88 & \\ This work &
  4.360 & 2.10 & 3.71 & \\ \hline Experiment & & & & \\ 3$C$ &
  4.360$^{d}$ & 2.24$^{e}$ & & \\ $*$ & & 2.234$^{f}$ & & \\ $*$ & &
  2.25$^{g}$ & & \\ 3$C$ & & 2.27$\pm$ 0.03$^{h}$ & 4.1 $\pm$
  0.10$^{h}$ & 0.425$^{h}$ \\ 3$C$ & &2.48$\pm$ 0.09$^{i}$ & 4.0 $\pm$
  0.3$^{i}$ & 0.25$^{i}$ \\ 6$H$,15$R$ & & 2.24$\pm$ 0.03$^{j}$ &
  4.3$\pm$ 0.3$^{j}$ & 0.45$^{j}$ \\ 3$C$,6$H$ & & 2.60$\pm$0.09$^{k}$
  & 2.9$\pm$0.3$^{k}$ & 0.95$^{k}$ \\
\end{tabular}
\end{center}
$^{*}$ Performed on specimens of $\alpha$-SiC (hexagonal and
rhombohedral phases).\\ $^{a}$ Ref.\cite{chang87}, $^{b}$
Ref.\cite{lambrecht91}, $^{c}$ Ref.\cite{karch94}, $^{d}$
Ref.\cite{landolt82}, $^{e}$ Estimated for 3$C$-SiC by Yean {\em et
  al.} \cite{yean71}. \\ $^{f}$ Schreiber {\em et al.}
\cite{schreiber66}. \\ $^{g}$ R. D. Carnahan \cite{carnahan68}. \\
$^{h}$ Aleksandrov {\em et al.} \cite{aleksandrov}. \\ $^{i}$
Str{\"o}ssner {\em et al.} \cite{strossner87}.\\ $^{j}$ Goncharov {\em
  et al.} \cite{goncharov90}.\\ $^{k}$ Yoshida {\em et al.}
\cite{yoshida93}.\\
\end{table}

\clearpage
\begin{table}
  \caption{Transition parameters from zincblende  structure to rocksalt
    structure for SiC.  $V_{t}$ is the volume of zincblende ($3C$)
    phase at transition, and $V_{0}$ is the volume at ambient
    pressure.  $\triangle V/V_{0}$ denotes the percentage of volume
    reduction when the transition occurs.  }
\label{trans-sic}
\vskip 8mm
\begin{center}
  \begin{tabular}{lccc}
    & Pressure (Mbar) & $V_{t}$/$V_{0}$ & $\triangle V/V_{0}$ \\
    \hline This work & 0.65 & 0.817 & 20.2\% \\ Chang {\it et
      al.}$^{a}$ & 0.66 & 0.81 & 18.5\% \\ Cheong {\it et al.}$^{b}$ &
    0.60 & 0.825 & \\ Christensen {\it et al.}$^{c}$ & 0.59 &
    0.84$^{d}$ & 19\%$^{d}$ \\ Experiment$^{e}$ & 1.00 & 0.757 &
    20.3\%\\
\end{tabular}
\end{center}
$^{a}$ Ref.\cite{chang87}.\\ $^{b}$ Ref.\cite{cheong91}.\\ $^{c}$
Ref.\cite{christensen87}.\\ $^{d}$ Quoted from Ref.\cite{yoshida93},
in which the authors derived these numbers from Fig.12 of
Ref.\cite{christensen87}.\\ $^{e}$ Ref.\cite{yoshida93}.\\
\end{table}

\clearpage
\begin{table}
\caption{Calculated Born effective charges $Z^{*}$ and $\epsilon_{\infty}$
  at experimental volume. }
\label{z-sic}
\begin{center}
\begin{tabular}{lcccc}
  & $Z^{*}$(Si) & $Z^{*}$(C) &$\frac{|Z^{*}(Si)|+|Z^{*}(C)|}{2} $ &
  $\epsilon_{\infty}$ \\ \hline This work & & & & \\ Number of special
  $k$-points & & & & \\ $~~~~~2$ & 2.428 & -3.392 & 2.910 & 9.426 \\
  $~~~~10$ & 2.701 & -2.716 & 2.708 & 7.137 \\ $~~~~28$ & 2.709 &
  -2.693 & 2.701 & 7.005 \\ \hline Karch {\it et al.}$^{a}$ & & & & \\
  (at LDA volume) & & & 2.72 & 6.97 \\ \hline Experiment & & &
  2.697$^{b}$ & 6.52$^{c}$ \\
\end{tabular}
\end{center}
$^{a}$ Ref.\cite{karch94}. \\ $^{b}$ Ref.\cite{olego82}. \\ $^{c}$
Ref.\cite{landolt82}. \\
\end{table}

\clearpage
\begin{table}
\caption{Comparison of phonon frequencies (${\rm cm}^{-1}$) in
  $3C$-SiC.}
\begin{tabular}{lcccccccccc}

  & $\Gamma_{\rm TO}$ & $\Gamma_{\rm LO}$ & $X_{\rm TA}$ & $X_{\rm
    LA}$ & $X_{\rm TO}$ & $X_{\rm LO}$ & $L_{\rm TA}$ & $L_{\rm LA}$ &
  $L_{\rm TO}$ & $L_{\rm LO}$\\ \tableline This work & 774 & 945 & 361
  & 622 & 741 & 807 & 257 & 601 & 747 & 817 \\ Karch {\it et
    al.}\tablenote{Reference \cite{karch94}.} & 783 & 956 & 366 & 629
  & 755 & 829 & 261 & 610 & 766 & 838 \\
  Experiment\tablenote{Reference \cite{feldman68}.} & 796 & 972 & 373
  & 640 & 761 & 829 & 266 & 610 & 766 & 838 \\
\end{tabular}
\label{freq}
\end{table}

\clearpage
\begin{table}
\caption{Elastic constants  and bulk modulus (in Mbar) of $3C$-SiC.}
\label{elastic-sic}
\vskip 8mm
\begin{center}
\begin{tabular}{lcccc}
  & $C_{44}$ & $C_{11}$ & $C_{12}$ & $B_{0}$ \\ \hline This work &
  2.41 & 3.84 & 1.32 & 2.16 \\ Karch {\it et al.}$^{a}$ & 2.53 & 3.90
  & 1.34 & 2.19 \\ Lambrecht {\it et al.}$^{b}$ & 2.87 & 4.20 & 1.26 &
  2.24 \\ Experiment$^{c}$ & 2.56 & 3.90 & 1.42 & 2.25 \\
\end{tabular}
\end{center}
$^{a}$ Ref.\cite{karch94}. \\ $^{b}$ Ref.\cite{lambrecht91} \\ $^{c}$
The experimental data are derived by Lambrecht {\it et al.}
\cite{lambrecht91} from the sound velocities of Feldman {\it et al. }
\cite{feldman68}.
\end{table}

\clearpage
\begin{table}
  \caption{ Gr\"uneisen parameters  of phonon modes in $3C$-SiC.}
\label{grun}
\vskip 8mm
\begin{center}
  \begin{tabular}{lcc}
    & This work & Expt.  \\ \hline TO($\Gamma$) & 1.07 & 1.02$^{a}$,
    1.102$^{b}$ \\ LO($\Gamma$) & 1.02 & 1.01$^a$, 1.091$^{b}$ \\
    \hline TA($X$) & 0.12 & \\ LA($X$) & 0.82 & \\ TO($X$) & 1.46 &
    1.30$^a$ \\ LO($X$) & 1.16 & \\ \hline TA($L$) & -0.13 & $-
    0.28^a$ \\ LA($L$) & 0.90 & $- 0.11^a$ \\ TO($L$) & 1.31 &
    1.24$^a$ \\ LO($L$) & 1.15 & 1.30$^a$ \\
\end{tabular}
\end{center}
$^{a}$ Converted from the results of Olego {\it et al.}
\cite{olego82,olego82-2}; see text. \\ $^{b}$ Aleksandrov {\it et al.
  } \cite{aleksandrov}.  \\
\end{table}

\clearpage

\begin{table}
\caption {Volume dependence of Born effective charge $Z$, dielectric constant
  $\varepsilon_\infty$, and LO-TO splitting in $3C$-SiC. The lattice
  parameter $a$ is in a.u. and frequencies $\omega$ are in cm$^{-1}$.}
\begin{tabular}{cccccc}

  $a$ & $Z^{*}$ & $\varepsilon_\infty$ & $\omega_{\rm TO}$ &
  $\omega_{\rm LO}$& $\omega_{\rm LO} - \omega_{\rm TO}$\\ \tableline
  \dec 8.339 & \dec 2.676 & \dec 7.111 & 744 & 910 & 166 \\ \dec 8.239
  & \dec 2.701 & \dec 7.005 & 775 & 945 & 171 \\ \dec 8.157 & \dec
  2.721 & \dec 6.927 & 800 & 975 & 175 \\ \dec 8.069 & \dec 2.744 &
  \dec 6.852 & 831 & 1010 & 179 \\ \dec 8.000 & \dec 2.761 & \dec
  6.800 & 855 & 1038 & 183 \\ \dec 7.864 & \dec 2.793 & \dec 6.707 &
  905 & 1094 & 189 \\ \dec 7.710 & \dec 2.830 & \dec 6.626 & 965 &
  1161 & 196 \\ \dec 7.610 & \dec 2.850 & \dec 6.575 & 1015 & 1214 &
  199 \\
\end{tabular}
\label{zvolt}
\end{table}

\begin{figure}
\caption{Calculated phonon frequencies of $3C$-SiC
  at the experimental lattice constant (solid lines).  The solid
  square symbols are from Raman measurements (see text).}
\label{dispersion}
\end{figure}

\begin{figure}
\caption{Pressure dependence of zone-center phonon frequencies of SiC.
  Filled circles are calculated results for 3$C$-SiC. Solid lines are
  the measurements for 3$C$-SiC up to 25 GPa (0.25 Mbar) by Olego $et$
  $al.$\protect\cite{olego82}.  Triangles are the experimental results
  of 6$H$-SiC reported by Liu and Vohra\protect\cite{liuvohra} to 95
  GPa (0.95 Mbar). }
\label{fre-lto-pressure}
\end{figure}

\begin{figure}
\caption{Pressure dependence of LO-TO splitting in  SiC.
  Filled circles connected with solid lines are the theoretical
  results for 3$C$-SiC using the calculated dielectric constants.
  Filled circles connected with dotted lines correspond to using the
  adjusted calculated dielectric constants for 3$C$-SiC (see text).
  The dashed line represents the measurement by Olego $et$
  $al.$\protect\cite{olego82} for 3$C$-SiC up to 25 GPa (0.25 Mbar).
  Triangles are the experimental results of 6$H$-SiC reported by Liu
  and Vohra\protect\cite{liuvohra} to 95 GPa (0.95 Mbar). }
\label{lto-split}
\end{figure}

\begin{figure}
\caption{
  Comparison of the scaling relation in Eq.  (\protect\ref{dedv}) to
  the calculated values (filled circles) of $\varepsilon_\infty$ for
  several values of $r$: (a) $r = 0.3$, (b) $r = 0.6$, and (c) $r =
  0.8$.}
\label{dedvfig}
\end{figure}

\begin{figure}
\caption{
  Volume dependence of the Born effective charge $Z^*$.  The filled
  circles connected by solid lines in ($a$) are the calculated $Z^*$
  for $3C$-SiC.  The dotted and dashed lines ($b-e$) are derived from
  Liu and Vohra's LO-TO splitting of 6$H$-SiC but using different
  equations of state and scaling relations for the volume dependence
  of dielectric constant (see text): ($b$) is determined using the
  scaling relation with $r=0.6$ and Yoshida {\it et al.}'s equation of
  state; ($c$) is determined using the scaling relation with $r=0.6$
  and our calculated equation of state; ($d$) is determined using
  thescaling relation with $r=0.3$ and Yoshida {\it et al.}'s equation
  of state; ($e$) is determined using the scaling relation with
  $r=0.3$ and our calculated equation of state.  }
\label{zvol}
\end{figure}

\end{document}